# Kinetic Economies

**Wan Ahmad Tajuddin Wan Abdullah\* and Sidiq Mohamad Khidzir**

Department of Physics, Universiti Malaya, 50603 Kuala Lumpur

We study a minimalist kinetic model for economies. A system of agents with local trading rules display emergent demand behaviour. We examine the resulting wealth distribution to look for non-thermal behaviour. We compare and contrast this model with other similar models.

\*Email: wat@um.edu.my

Fax: +603-79674146

# I. INTRODUCTION

A pressing problem to physicists trying to understand economic systems is precisely just that – i.e. what is an economy? This is in the sense of statistical physics – we would like to know what makes an economic system different (if it is!) from the kinetic view of gases, for example. What is the essence of a system that would make it an 'economic' system? Can it be treated with the current arsenal of statistical physics, or does this difference or essence require something more?

We attempt at this problem by looking for minimalist models that may show economic behaviour. We take a bottom-up approach, constructing kinetic systems of 'agents' with local interactions (trading rules), and look at emergent system behaviour. Subsequently interactions may be increasingly complexified to produce what may be required as economic behaviour.

# II. THE MODEL

As reported earlier [1], the model consists of agents $i$, each with amount of tradable goods $b_i$, money $d_i$, and perception of price $h_i$, randomly 'hitting' each other in pairs, doing trade as they do. The trading rules are given in Fig. 1 where we have chosen a buyer's model – you buy if the offered price is lower, and you buy at whatever price if you have no goods. Notice that the interactions are conservative in money and goods but not in price. There is also scale independence of money and goods and the model is adequately described by the overall ratio of goods to money. The interactions are local – there is no global common knowledge available to agents, and are zero-temperature – there is no noise.

We have already shown that this model yields a demand-like curve in the price as we vary the amount of goods available, reminiscent of economic systems. The 'greed' or profit-maximizaing behaviour, expected to be underlying to produce such behaviour, is actually intrinsic in the interactions. Though price information is local, we expect it to equilibrate through the meeting of agents. Figs. 2 show the evolution of the price distribution. The dynamics are dissipative (in *h*) – there is tendency to move to the lowest state, but this is counteracted by a 'pump' – zero *b* states are pushed to higher *h*.

## III. WEALTH DISTRIBUTION

It has been shown that [2,3] the wealth distributions of people in various countries generally consist of two portions – the Boltzmann-Gibbs compliant lower one, and a Paretto power-law upper tail. The masses accumulate wealth in a thermal random-walk fashion while the wealthier class do it differently, explained [2] as respectively with additive and with multiplicative income processes.

Wealth distributions for models with money exchange [4], money exchange with saving propensity [5] and asymmetric money exchange [6] have been reported. As argued [4], models with conservation of money like these, yield Boltzmann-Gibbs behaviour as do kinetic models with conservation of energy. Of course, this is assuming ergodicity; non-thermal behaviour can arise in the absence of ergodicity [7]. Interestingly, a stock market simulation game (http://us.newsfutures.com/) seems to yield the Paretto type behaviour as in the upper portion.

Fig. 3 shows the wealth distribution obtained from our model, where wealth is *d+hb*, for different *d* to *b* ratios. For a 1:1 ratio, Fig. 4a depicts a log-linear plot while Fig. 4b shows the log-log form. The region of wealth starting from just below 50 up to almost 250 portrays thermal behaviour, while just beyond that a power law dominates. The Paretto region is however is very steep and thins the thermal tail rather than fattens it, due perhaps to the dissipative nature of *h*. The region of wealth below the thermal region shows interesting behaviour, perhaps resulting from the 'pumping' of *h*.

As *h* is not conserved, so is wealth not conserved in the model. Also, as values for *h* are fixed at the start, it is not ergodic. These provide rationale for the non-thermal behaviour.

## IV. CONCLUSION

In summary, we simulated a minimal model in an attempt to look for the essence of an economy, which displayed thermal as well as non-thermal behaviour. We may extend this to be more symmetric by including the seller model, or by having a more rational rule for price which may be a scaled ratio between the offers of the buyer and the seller. Further, inclusion of temperature, multiple goods, production and consumption, etc may also be explored.

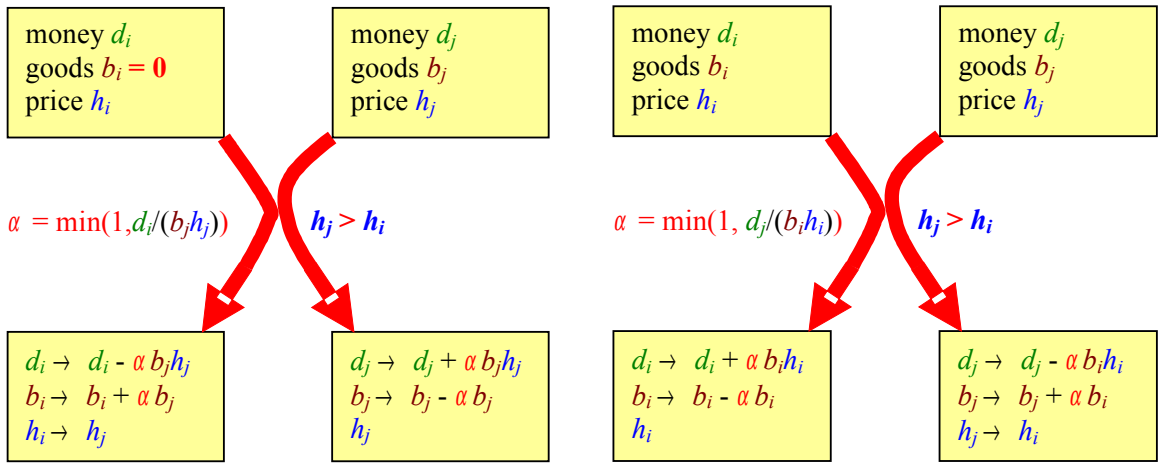

Fig. 1. Trading rules.

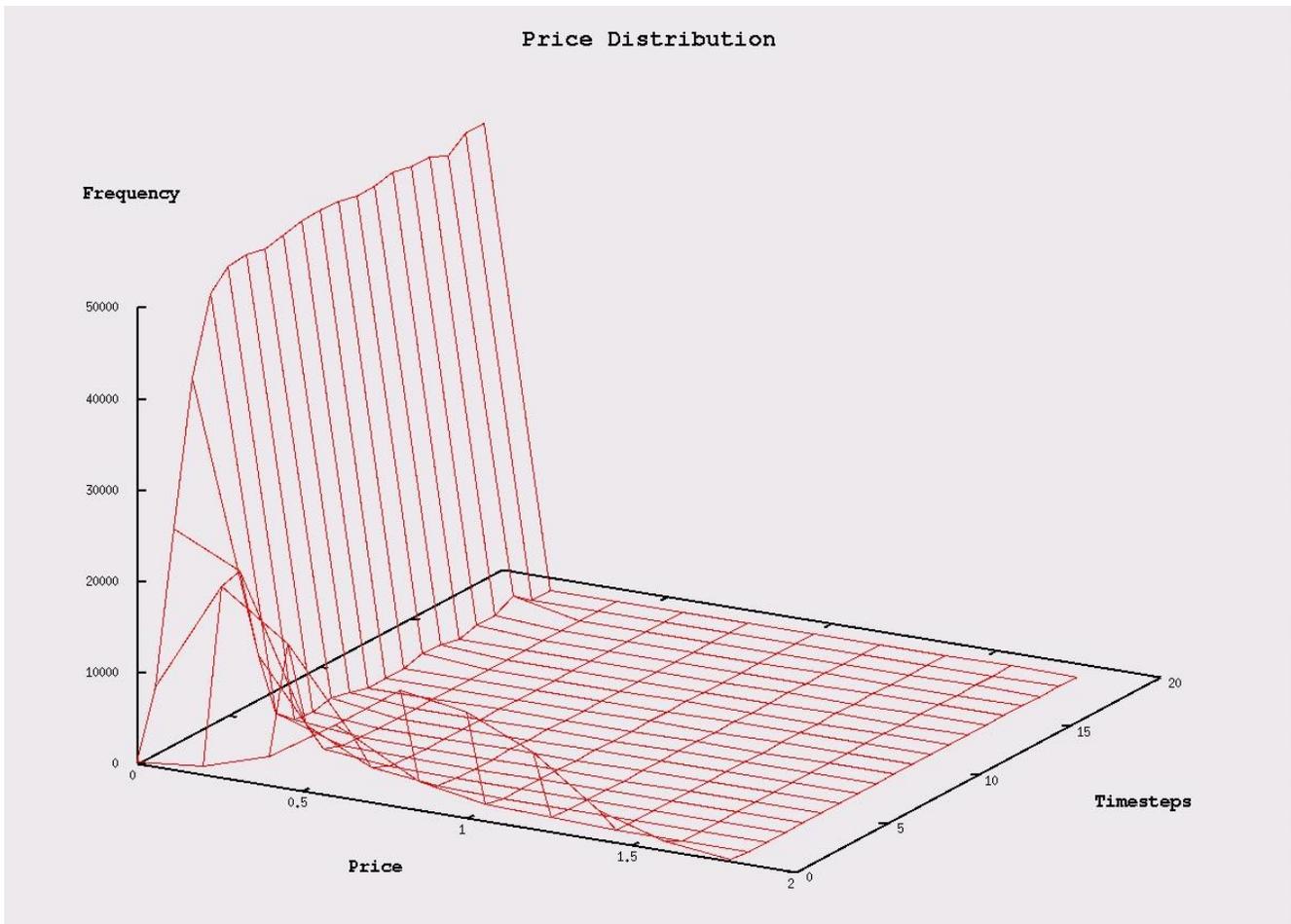

Fig. 2a. Evolution of the price distribution, for a 1:1 money to goods ratio.

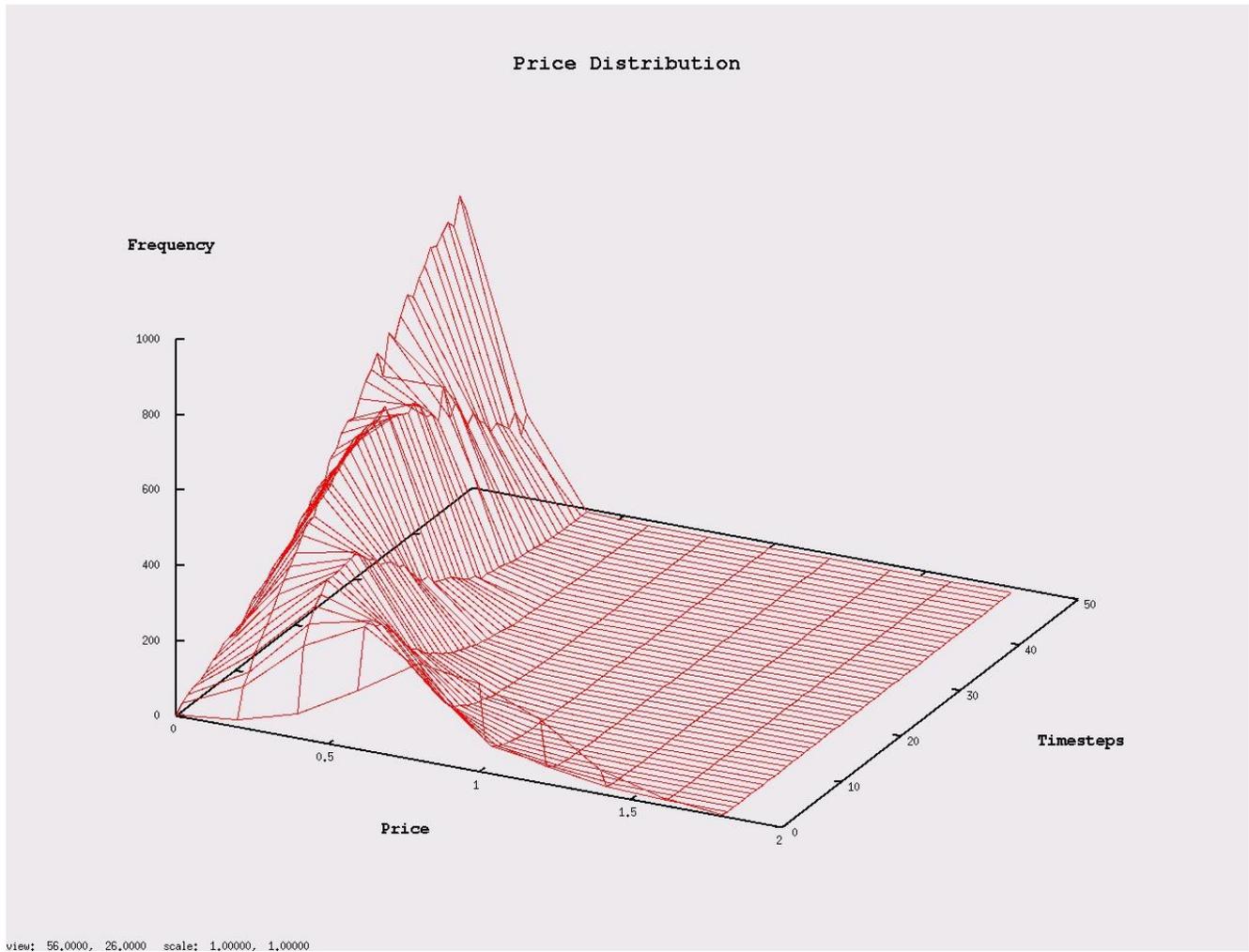

Fig. 2b. Evolution of the price distribution, for a 100:1 money to goods ratio.

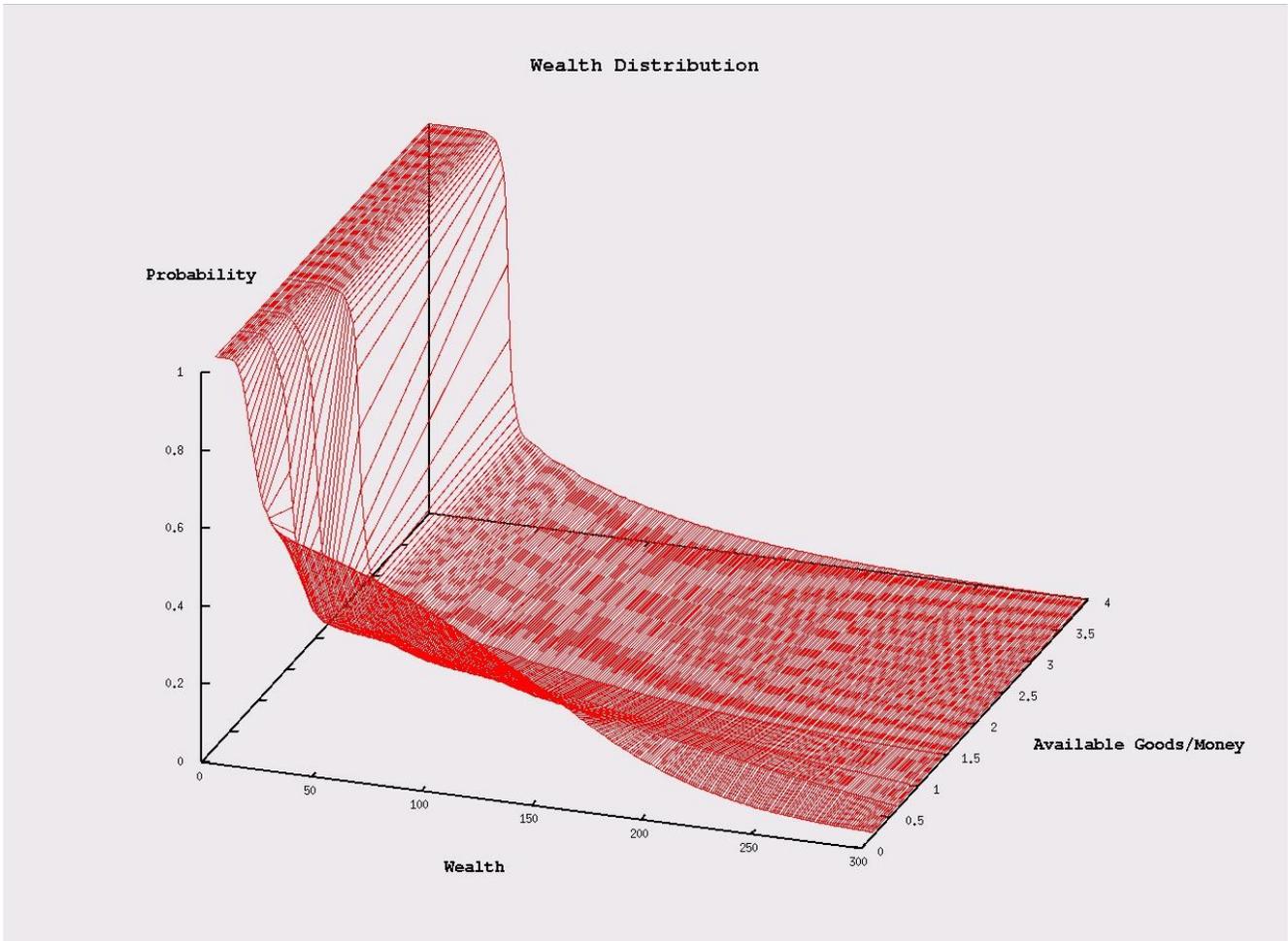

Fig. 3. (Cumulative) wealth distribution for different goods/money ratios.

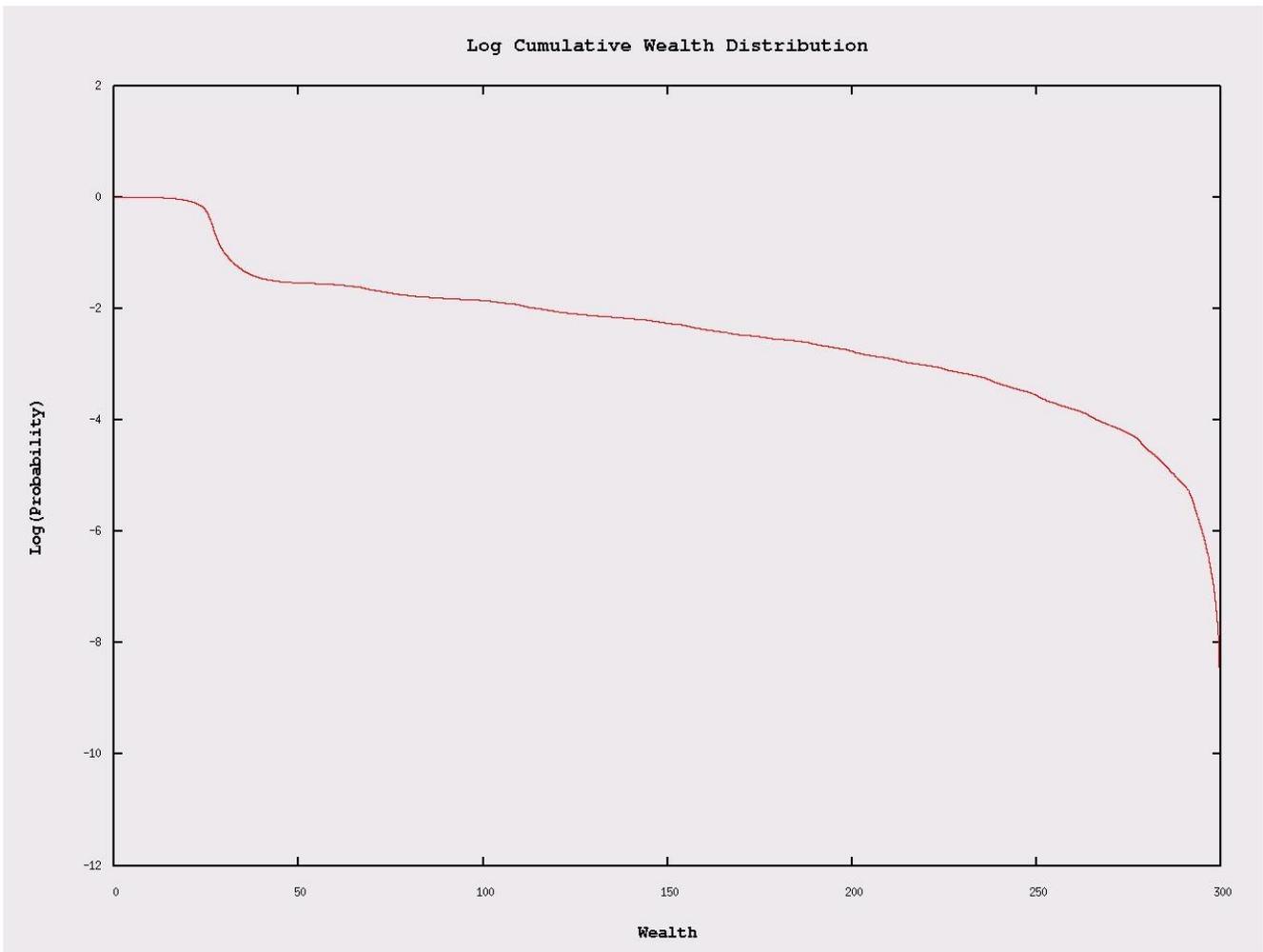

Fig. 4a. Log-linear display of wealth distribution for a 1:1 goods to money ratio

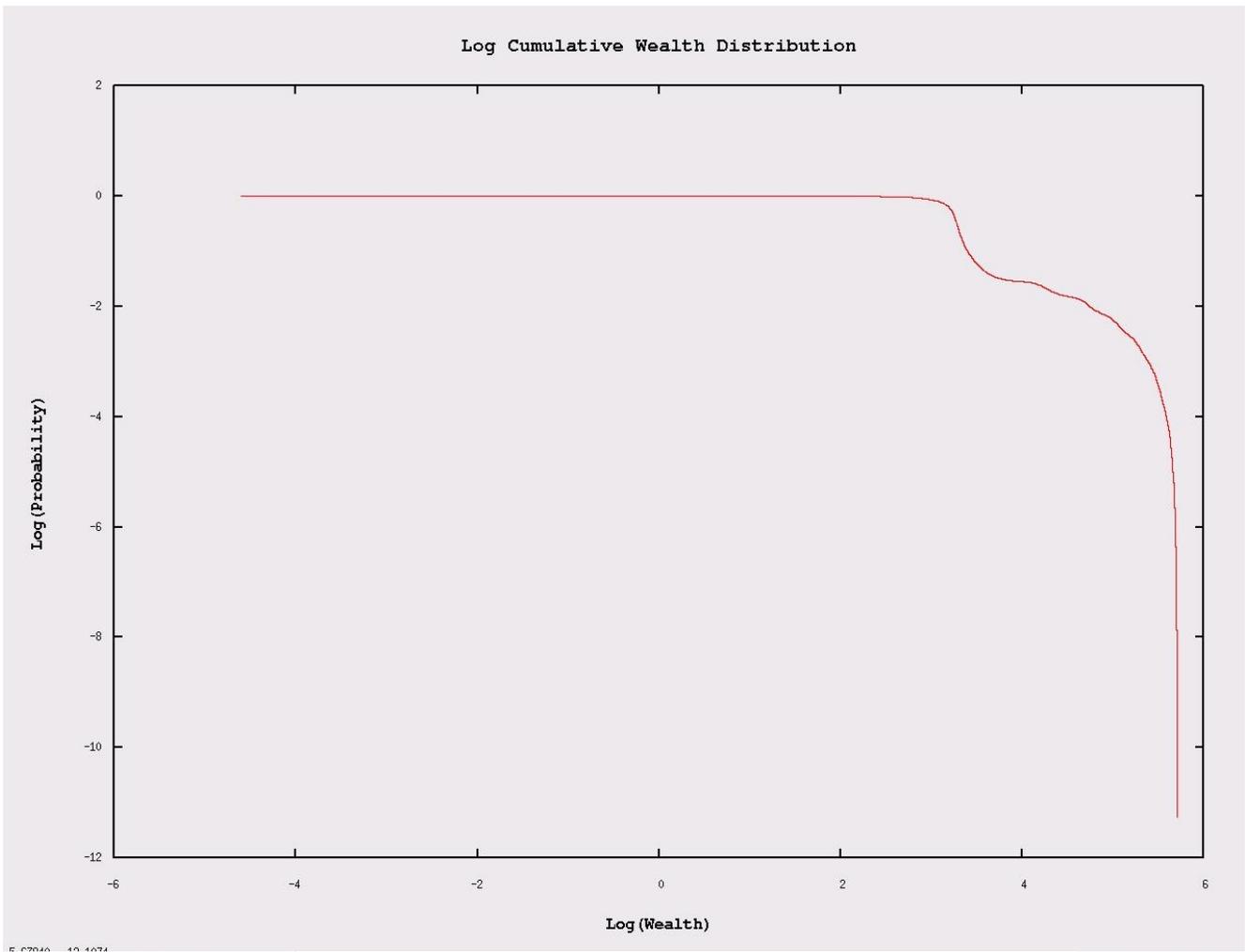

Fig. 4a. Log-log display of wealth distribution for a 1:1 goods to money ratio